\documentstyle[aps,amsfonts,epsfig]{revtex}
{\makeatletter%
\gdef\labeleqs#1{{%
\edef\@currentlabel{%
\ifappendixon\appletter\fi
\ifsecnumbers\ifnum\c@secnum>0
\arabic{secnum}.\fi\fi\arabic{equation}}%
\label{#1}%
}}%
}

\twocolumn
\begin{document}
\newcommand{\beq}{\begin{eqnarray}}
\newcommand{\eeq}{\end{eqnarray}}
\newcommand{\diff}{\mbox{d}}
\newcommand{\pardis}{\langle \mu \rangle}
\newcommand{\mathz}{\mathbb{Z}}
\newcommand{\proj}{\vec{\Phi}(x)}
\newcommand{\vproj}{\hat{\Phi}(x)}
\newcommand{\latproj}{\hat{\Phi}(\vec{n}+\hat{\imath},t)}
\newcommand{\latbproj}{\hat{\Phi}(\vec{n},t)}
\draft
\widetext
\title{Colour confinement and dual superconductivity of the vacuum - I}

\author{A. Di Giacomo$^{a,c,1}$, B. Lucini$^{b,c,2}$,
L. Montesi$^{b,c,2}$, G. Paffuti$^{a,c,1}$}

\address{$^a$ Dipartimento di Fisica dell'Universit\`a, Via Buonarroti
2 Ed. B, I-56127 Pisa, Italy}

\address{$^b$ Scuola Normale Superiore, Piazza dei Cavalieri 7, I-56126
Pisa, Italy}

\address{$^c$ INFN sezione di Pisa, Via Vecchia Livornese 1291, I-56010
S. Piero a Grado (Pi), Italy}

\address{$^1$ e-mail address: digiaco, paffuti@mailbox.difi.unipi.it}
\address{$^2$ e-mail address: lucini, montesi@cibs.sns.it}

\maketitle
\widetext
\begin{abstract}
We study dual superconductivity of the ground state of $SU(2)$
gauge theory, in connection with confinement. We do that measuring on the
lattice a disorder parameter describing condensation of monopoles. Confinement
appears as a transition to dual superconductor, independent of the
abelian projection defining monopoles. Some speculations are made
on the existence of a more appropriate disorder parameter. A similar
study for $SU(3)$ is presented in a companion paper.
\end{abstract}
\pacs{PACS numbers: 11.15.Ha, 12.38.Aw, 14.80.Hv, 64.60.Cn}
\setcounter{page}{1}
\narrowtext
\section{Introduction}
Order-disorder duality \cite{Kramers,Kadanoff} plays an increasingly important
r\^ole in our understanding of the dynamics of gauge theories, specifically
of {\em QCD} \cite{'tHooft2,Digia1} and of its supersymmetric generalizations
\cite{Seiberg}.

Duality is typical of systems which can have configurations with non
trivial spatial topology, carrying a conserved topological charge.
The prototype example is the $2d$ Ising model. If viewed as a discretized
version of a $1+1$ dimensional field theory, it presents one-dimensional
configurations, kinks, whose topology is determined by the boundary conditions
($\pm 1$) at $x_1=\pm \infty$.

In the usual description in terms of the local variable $\sigma(x) = \pm 1$,
at low temperature (weak coupling) the system is in an ordered phase with
nonzero magnetization $\langle \sigma \rangle \ne 0$. At the critical
point $\langle \sigma \rangle \to 0$ and the system becomes disordered.
$\langle \sigma \rangle$ is called an order parameter. However one can
describe the system in terms of a dual variable $\sigma ^{\star}$, on a dual
lattice. A dual $1d$ configuration with one $\sigma ^{\star}$
up is a kink, which is a highly non local object in terms of $\sigma$.
In ref. \cite{Kadanoff} it was shown that the partition function in terms of
$\sigma ^{\star}$ has the same form as in terms of $\sigma$, i.e.
that the system with dual description looks again as an Ising model,
except that the new Boltzmann factor $\beta ^{\star}$ is related to the old
one $\beta$ by the relation
\begin{eqnarray}
\sinh \left( 2 \beta ^{\star}\right) = \frac{1}{\sinh \left( 2 \beta
\right)} \ .
\end{eqnarray}
The disordered phase is an ordered phase for the dual and viceversa.
In the disordered phase $\langle \sigma ^{\star} \rangle \ne 0$: kinks
condense in the ground state. $\langle \sigma ^{\star} \rangle$ is
called a disorder parameter.
$\sigma ^{\star}$ is a dual variable to $\sigma$. In this specific case
the system is self dual, and duality transformation maps the strong coupling
regime in the weak coupling regime and viceversa.\\

~\vphantom{a}
\vskip7\baselineskip\noindent
Other systems showing duality properties are the $3d$
$XY$ model, whose dual is a Coulomb gas in $3d$, and the compact $U(1)$ gauge
theory.

In the $3d$ $XY$ model topological excitations are the vortices of the $2d$
$XY$ model. These vortices condense in the disordered phase \cite{CE97}.

For $U(1)$ theory topological excitations are monopoles.
There the duality transformation can be performed for special choices of the
action (e.g. the Villain action \cite{Marchetti}, dual to a $\mathz$ gauge
theory, and the Wilson action \cite{Vinc}). For other choices it is not
known how to explicitly perform the transformation to the dual.

An alternative approach consists in identifying the symmetry which is
spontaneously broken in the disordered phase, i.e. the topological
configurations which are supposed to condense, and in writing a disorder
parameter in terms of the original local fields \cite{Swieca}.
The disorder parameter is then the vacuum expectation value ({\em vev})
of a non local operator.

This approach has been translated on the lattice \cite{Digiacta,Deldebbio},
tested by numerical simulations in the compact $U(1)$
gauge theory \cite{Digiau1}, in the $3d$ $XY$ model \cite{CE97} and in the
$O(3)$ sigma model \cite{Martelli}, and first used to investigate colour
confinement in {\em QCD} in ref. \cite{Pieri}.

In the early literature on the subject condensation was demonstrated
as the sudden increase of the density of topological excitations. This
is incorrect, since disorder can only be described by the {\em vev}
of an operator which violates the dual symmetry and the number of excitations
does not.

Looking at symmetry is specially important in {\em QCD}. For {\em QCD}
there exists some
general idea about the dual \cite{'tHooft2,Polyakov}.
The dual description should
also be a gauge theory, possibly with interchange of the r{\^o}le of electric
and magnetic quantities.

This idea could fit the mechanism for confinement of colour proposed in ref.'s
\cite{Mandel,'tHooft4} as dual superconductivity of the ground state, if
confinement were due to disorder and monopoles were the topological excitations
which condense.
However a dual superconductor is a typically abelian system, while the
disorder parameter is expected to break a non abelian symmetry. An abelian
conserved monopole charge can be associated to each operator in the adjoint
representation by a procedure which is known as abelian projection
\cite{'tHooft3}. We will recall that procedure in Sect.~\ref{construction}.
There exists a functional infinity of choices for the operator, and
correspondingly an infinity of monopole species. A possibility is that
the true disorder symmetry implies the condensation of all these species of
monopoles \cite{'tHooft3}.
Some people believe instead
that some abelian projection (specifically the Maximal
Abelian one) identifies monopoles that are more relevant than others
for confinement.
Both attitudes reflect our ignorance of the dual description of the system.

In this paper we shall systematically explore condensation of monopoles
defined by different abelian projections, in connection with confinement of
colour.

We will do that for $SU(2)$ gauge theory. The treatment of $SU(3)$ will be
given in a companion paper. Some of the results have been obtained
during the last years and have been reported to conferences and
workshops \cite{Digiacta,Storia1,Storia2}.
This paper contains conclusive results, and is an organic report of the
methods and of the results obtained after ref. \cite{Pieri}.

Our strategy consists in constructing an operator with non zero magnetic
charge, for each abelian projection (Sect.~\ref{disorder}).
Its {\em vev} is a candidate disorder parameter
for dual superconductivity of the ground state.
We shall determine numerically that {\em vev} at finite
temperature below and above the deconfining phase transition. If condensation
of these monopoles is related to confinement, we expect the disorder parameter
to be zero in the deconfined phase, and different from zero in the confined
phase.

This is strictly speaking true only in the thermodynamic (infinite volume)
limit. A finite size scaling analysis allows to go to that limit, and, as a
by-product, gives a determination of the transition temperature and of the
critical indices if the transition is higher order than first. This analysis
is presented in Sect.~\ref{results}.

A special treatment for the Maximal Abelian projection is presented in
Sect.~\ref{maxabsection}.

We find that gauge theory vacuum is indeed a dual superconductor in the
confined phase, and becomes normal in the deconfined phase for a number
of abelian projections, actually for all projections that we have analyzed,
in agreement with the guess of ref. \cite{'tHooft3}.

The idea that confinement is produced by dual superconductivity is thus
definitely confirmed. The guess that all the abelian projections are
physically equivalent is also supported, and this is an important piece of
information on the way to understand the true dual symmetry.

We find evidence that $SU(2)$ deconfining transition is second order.
In next paper we will show that for $SU(3)$ this transition is first order.

An analysis of full {\em QCD}, including quarks, is on the way; if the mechanism
proved to be the same, the idea that quarks are a kind of perturbation,
and that the dynamics is determined by gluons would be tested.
This would also be a test of the ansatz that the theory already contains
its essential dynamics at $N_c = \infty$, and that the presence of
fermions and the extrapolation to $N_c = 3$ can be viewed as perturbations.

The results are summarized in Sect.~\ref{conclusions}.
\section{The abelian projection}
\label{construction}
What follows will refer to the case of gauge group $SU(2)$. Adaptation
to $SU(3)$ will be described in the companion paper.

Let $\vproj$ be the direction in colour space of any local operator
$\proj$, belonging to the adjoint representation of $SU(2)$. A gauge
transformation $g(x)$ which rotates $\vproj$ to $(0,0,1)$, or which
diagonalizes $\vproj \cdot \vec{\sigma}$ is called the abelian projection
on $\proj$. $g(x)$ can be singular in a configuration at the points
where $\proj$ has zeros, and $\vproj$ is not defined.

The field strength $F_{\mu \nu}$, defined as \cite{'tHooft1}
\beq
\label{eq:2.1}
F_{\mu \nu} = \vproj \cdot \vec{G}_{\mu \nu} - \frac{1}{g}
\left( D_{\mu} \vproj \wedge D_{\nu} \vproj \right) \cdot \vproj
\eeq
is a colour singlet, and is invariant under non singular gauge transformations.

In general eq. (\ref{eq:2.1}) can be written as \cite{Arafum}
\beq
\label{eq:2.2}
F_{\mu \nu} &=&  {\partial}_{\mu} \tilde{A}_{\nu} - {\partial}_{\nu}
\tilde{A}_{\mu}\\
\nonumber
&-&\frac{1}{g} \left( \partial _{\mu} \vproj \wedge \partial _{\nu} \vproj
\right) \cdot \vproj \ .
\eeq
with
\begin{eqnarray}
\tilde{A}_{\mu} = \hat{\Phi}^a A_{\mu}^a \ .
\end{eqnarray}
In the abelian projected gauge $\vproj$ is constant, the second term in the
right hand side of eq.~(\ref{eq:2.2}) vanishes and the field $F_{\mu \nu}$
becomes an abelian field.

Denoting by $F^{\star}_{\mu \nu}$ the usual dual tensor,
\beq
\label{eq:2.3}
F^{\star}_{\mu \nu} = \frac{1}{2} \epsilon _{\mu \nu \rho \sigma} F^{\rho
\sigma} \ ,
\eeq
and defining the magnetic current as
\beq
\label{eq:2.4}
j_{\mu} &=& {\partial}^{\nu} {F}^{\star}_{\mu \nu} \ ,
\eeq
it follows from eq.'s~(\ref{eq:2.2}),~(\ref{eq:2.3}),~(\ref{eq:2.4})
that
\beq
\label{eq:2.5}
\partial^{\mu} j_{\mu} = 0 \ .
\eeq
The magnetic charge is conserved, and defines a magnetic $U(1)$ symmetry.

The abelian projection $g(x)$ can have singularities and as a consequence an
additional field strength adds to the usual covariant gauge transform
of $G_{\mu \nu}$ \cite{'tHooft1}.

After abelian projection
\beq
G_{\mu \nu} = g G_{\mu \nu} g^{-1} + G_{\mu \nu} ^{sing} \ ,
\eeq
with $\vec{G} _{\mu \nu} ^{sing} = \vproj \left( \partial _{\mu}
\tilde{A}_{\nu} ^{sing} - \partial_{\nu} \tilde{A}_{\mu} ^{sing} \right)$
parallel to the colour direction $\vproj$, and consisting of Dirac strings
starting at the zeros of $\proj$. The field configurations contain
monopoles at the zeros of $\proj$, as sinks or sources of the regular
field, and the strings carry away the corresponding magnetic flux.

On a lattice (or in any other compact regularized description in terms
of parallel transport) the Dirac string reduces to an additional flux of
$2 \pi$ across a sequence of plaquettes, which is invisible \cite{Degrand}.

The mechanism relating confinement of colour to dual superconductivity of
the vacuum advocates a spontaneous breaking {\em \`a la Higgs} of the
magnetic $U(1)$ symmetry described by (\ref{eq:2.5}), which constrains
the electric component of the field eq. (\ref{eq:2.2}) into flux tubes.

All particles which have non zero electric charge with respect to the residual
$U(1)$ (\ref{eq:2.2}) will then be confined. There exist coloured
states, e.g. the gluon oriented parallel to $\vproj$ which are not confined.
This is a strong argument that, if dual superconductivity is a mechanism of
confinement at all, it must exist in many different abelian projections,
as a manifestation of non abelian disorder.

On the lattice the abelian gauge field corresponding to any given projection,
or $\vproj$, is extracted as follows \cite{Kronfeld}.

Let $\bar{U}_{\mu}(n)=g(n) U_{\mu}(n) g^{\dagger}(n)$ be the generic link
after abelian projection. We adopt the usual notation
$U_{\mu}(n) \equiv U_{\mu}(\vec{n},t) \equiv \exp(i A_{\mu}(n))$, with
$A_{\mu}(n)= \vec{A}_{\mu}(n) \cdot \vec{\sigma}$.

The representation in terms of Euler angles has the form \beq
\bar{U}_{\mu} &=& e^{i \alpha_{\mu} \sigma _3} e^{i \gamma_{\mu}
\sigma_2} e^{i \beta _{\mu} \sigma_3} \label{eulero}
\\ \nonumber
&=& e^{i \alpha_{\mu} \sigma _3} e^{i \gamma_{\mu} \sigma_2} e^{-
i \alpha_{\mu} \sigma _3} e^{i
\left(\beta_{\mu}+\alpha_{\mu}\right) \sigma_3}\\ \nonumber &=&
e^{i \vec{\gamma}_{\mu}^{T} \cdot \vec{\sigma}} e^{i \theta _{\mu}
\sigma_3} \ , \qquad \theta_{\mu} = \alpha _{\mu} + \beta _{\mu} \
, \eeq and $\vec{\gamma}^{T}$ is a vector perpendicular to the $3$
axis. We assume the usual representation in which $\sigma_3$ is
diagonal.

For a plaquette, a similar decomposition can be performed,
\beq
\label{eq:3.10}
\Pi _{\mu \nu} = e^{i \vec{\gamma}^T_{\mu \nu} \cdot \vec{\sigma}}
e^{i \theta _{\mu \nu} \sigma _3} \ ,
\eeq
$\theta _{\mu \nu} = \triangle _{\mu} \theta_{\nu} - \triangle _{\nu}
\theta _{\mu}$ up to terms ${\cal O}(a^2)$. $\theta _{\mu \nu}$ is the lattice
analog of $F_{\mu \nu}$. The abelian magnetic flux is
conserved by construction. A monopole appears whenever the flux entering
five faces of a spatial cube adds to more than $2 \pi$:
then the flux through the sixth face is larger than $2 \pi$, but
multiples of $2 \pi$ are invisible in the exponent. Formally \cite{Degrand}
\beq
\theta _{\mu \nu} = \bar{\theta}_{\mu \nu} + 2 \pi n_{\mu \nu} \ ,
\eeq
with $- \pi < \bar{\theta} _{\mu \nu} < \pi$. A string through the sixth face
takes care of the flux which has disappeared.

We shall construct a disorder parameter for monopole condensation as the
{\em vev} of an operator carrying non zero magnetic charge,
$\mu$. $\pardis \ne 0$ will signal dual superconductivity.
\section{The disorder parameter}
\label{disorder}
The disorder parameter will be constructed on the same lines as in ref.'s
\cite{CE97,Digiau1}.

An improvement exists with respect to ref. \cite{Pieri}, which consists
in properly taking the compactness into account: in ref. \cite{Pieri} the
approximation was that the field was treated as non compact. The same
improvement was done in ref. \cite{Digiau1} with respect to
ref.~\cite{Deldebbio}.

All the results presented in ref.'s
\cite{Storia2} already contain such improvement.

We first analyze the case in which $\proj$ is determined by the Wilson
Polyakov line, i.e. the closed parallel transport to $+ \infty$ along the
time axis and back from $- \infty$ to the initial point via the periodic
boundary conditions.
For this choice, after abelian projection all the links $U_0(n)$
along the temporal axis are diagonal, of the form $\bar{U}_0(n) \equiv
\exp(i A^3_{0}(n) \sigma _3)$.

Assuming for sake of definiteness the Wilson action we construct the operator
$\mu(\vec{y},t)$ which creates a monopole at site $\vec{y}$ and time $t$ with
the following recipe (a similar construction can be made for other action).

Let $\vec{A}^M(\vec{x},\vec{y})$ be the vector potential describing
the field value at site $\vec{x}$ of a static monopole sitting at $\vec{y}$.
We shall write it as
\beq
\vec{A}^M(\vec{x},\vec{y}) = \vec{A}^M_{\bot} (\vec{x},\vec{y}) +
\vec{\nabla} \Lambda (\vec{x},\vec{y}) \ ,
\eeq
with $\vec{\nabla} \cdot \vec{A}^M_{\bot} (\vec{x},\vec{y}) = 0$.

The first term describes the physical part of $\vec{A}^M$, the second
term the classical gauge freedom.

Let $\Pi _{i0}$ be the electric field plaquette at time $t$. Then
we define
\beq
\label{3:12}
\mu = \exp \left[ - \beta \Delta S \right] \ ,
\eeq
\beq
\label{eq:3.13}
\Delta S = \frac{1}{2} \sum _{\vec{n}}
\mbox{Tr} \left\{ \Pi _{i0}(\vec{n},t) - \Pi _{i0} ^{\prime} (\vec{n},t)
\right\} \ .
\eeq
Here
\beq
\Pi _{i 0} (\vec{n},t) &=& U_i(\vec{n},t)U_0(\vec{n}+ \hat \imath,t)\\
\nonumber
& &\left(U_i (\vec{n},t+1)\right)^{\dag}\left(U_0(\vec{n},t)\right)^{\dag}
\eeq
is the electric field term of the action, and $\Pi^{\prime} _{i 0}$
is a modification of it, defined as
\beq
\label{eq:3.15}
\Pi _{i 0} (\vec{n},t) ^{\prime}
&=& U_i(\vec{n},t)U_0(\vec{n}+ \hat \imath,t)\\
\nonumber
&&\left(U_i ^{\prime}(\vec{n},t+1)\right)^{\dag}
\left(U_0(\vec{n},t)\right)^{\dag} \ ,
\eeq
\beq
\label{eq:3.16}
U_i^{\prime}(\vec{n},t+1) &=& e^{i \Lambda (\vec{n},\vec{y}) \latbproj \cdot
\vec{\sigma}}U_i(\vec{n},t) \\
\nonumber
&&e^{-i A_{\bot i}^M (\vec{n},\vec{y}) \latproj \cdot
\vec{\sigma}}
e^{-i \Lambda (\vec{n}+\hat{\imath},\vec{y}) \latproj \cdot
\vec{\sigma}} \ .
\eeq

The disorder parameter is defined as $\pardis$, or
\beq
\label{eq:3.17}
\pardis = \frac{\int \left({\cal D} U \right) e^{- \beta (S + \Delta S)}}
{\int \left({\cal D} U \right) e^{- \beta S}} \ .
\eeq
It follows from the definition
(\ref{eq:3.13}) that adding $\Delta S$ to the action
amounts to replace the term $\Pi _{i 0}$ at time $t$ with
$\Pi _{i 0}^{\prime}$.

The $\Pi_{i0} (\vec{n},t)$ are the only terms in the action where the
$U_0(\vec{n},t)$ appear. In the path integral (\ref{eq:3.17})
a change of variables $U_0(\vec{n},t) \to U^{\prime}_0(\vec{n},t)=
U_0(\vec{n},t)e^{i \Lambda (\vec{n},\vec{y}) \hat{\Phi}(\vec{n},t) \cdot
\vec{\sigma}}$ leaves the Haar measure
invariant and reabsorbs the unphysical gauge factor of eq. (\ref{eq:3.16}),
so that $\pardis$ is independent, as it must be, of the choice of the
classical gauge for the field produced by the monopole.

Also a change of variables can be made
\beq
\label{eq:3.18}
U_i(\vec{n},t+1) \to U_i
(\vec{n},t+1)e^{i A_{\bot i}^M (\vec{n},\vec{y}) \latproj \cdot
\vec{\sigma}} \ .
\eeq

Again, this leaves the measure invariant, and brings $\Pi _{i 0}(\vec{n},t)$
to its original form. However in the plaquette $\Pi _{i j}(\vec{n},t+1)$
it produces the change $U_i(\vec{n},t+1) \to U_i(\vec{n},t+1)
e^{i A_{\bot i}^M (\vec{n},\vec{y}) \latproj \cdot \vec{\sigma}}$.
By the construction of Sect.~\ref{construction} this amounts to change,
in the abelian projected gauge
\beq
\label{eq:3.19}
\theta _{i j} (\vec{n},t+1) &\to& \theta _{i j} (\vec{n},t+1) +
\\
\nonumber
&&\triangle _i
A_{\bot j}^M (\vec{n},\vec{y}) -  \triangle _j A_{\bot i}^M (\vec{n}, \vec{y})
\eeq
or to add the magnetic field of a monopole.

The same redefinition of variables reflects in the change
\beq
\label{eq:3.20}
\Pi _{i 0} (\vec{n},t+1) \to \Pi ^{\prime} _{i 0} (\vec{n},t+1) \ ,
\eeq
analogous to equation (\ref{eq:3.15}). Again the gauge factors
$e^{- i \Lambda(\vec{n},\vec{y}) \hat{\Phi}(\vec{n},t) \cdot
\vec{\sigma}}$,
$e^{ i \Lambda(\vec{n} + \hat{\imath},\vec{y}) \latproj \cdot \vec{\sigma}}$
are irrelevant, since they can be reabsorbed
in a redefinition of $U_0(\vec{n},t+1)$ . $e^{i A_{\bot i}^M(\vec{n},\vec{y})
\latproj \cdot \vec{\sigma}}$ commutes with $U_0(\vec{n}+\hat{\imath},t+1)$,
which is diagonal with it by definition of the Polyakov line abelian
projection.

In detail
\beq
\Pi _{i 0}  ^{\prime} (\vec{n},t+1) &=&
U_i(\vec{n},t+1) e^{i A_{\bot i}^M(\vec{n},\vec{y}) \latproj \cdot
\vec{\sigma}}\\
\nonumber
&&
U_0(\vec{n}+ \hat \imath,t+1)\\
\nonumber
& &
\left(U_i (\vec{n},t+2)\right)^{\dag}\left(U_0(\vec{n},t+1)\right)^{\dag} \\
\nonumber
&=&  U_i(\vec{n},t+1) U_0(\vec{n}+ \hat \imath,t+1)\\
\nonumber
&&e^{i A_{\bot i}^M(\vec{n},\vec{y}) \latproj \cdot \vec{\sigma}}\\
\nonumber
&\ &
\left(U_i (\vec{n},t+2)\right)^{\dag}\left(U_0(\vec{n},t+1)\right)^{\dag} \ .
\eeq
A new change of variable can be done
analogous to (\ref{eq:3.18}), exposing now a monopole at $t+2$ and
producing a change $\Pi _{i0} (\vec{n},t+2) \to \Pi
^{\prime} _{i0} (\vec{n},t+2)$. The procedure can be iterated.
If an antimonopole is created at $t+T$, by an operator
analogous to that of (\ref{3:12}), but with
$\vec{A}_{\bot}^M \to - \vec{A}_{\bot}^M$,
then at time $t+T$ the change cancels and the procedure stops.

This shows that the correlation function
\beq
{\cal D}(T) = \langle \bar{\mu} (\vec{y},t+T) \mu (\vec{y},t) \rangle
\label{corr1}
\eeq
indeed describes the creation of a monopole at $\vec{y}$ at time $t$ and
its propagation from $t$ to $t+T$.
This argument in this gauge is perfectly analogous to the argument for compact
$U(1)$ gauge theory \cite{Digiau1}. The construction is the compact version
of that of ref. \cite{Pieri}.

At large $T$, by cluster property
\beq
{\cal D}(T) \simeq A \exp( - M T) + \pardis ^2 \ ,
\eeq
where the equality $\pardis = \langle \bar{\mu} \rangle$ has been used stemming
from charge conjugation invariance.

$\pardis \ne 0$ indicates spontaneous breaking of the $U(1)$ magnetic symmetry
defined in Sect.~\ref{construction} eq. (\ref{eq:2.5}), and hence dual
superconductivity \cite{Storia2}.
$\pardis$ is the corresponding disorder parameter.
In the thermodynamic limit we expect $\pardis \ne 0$ below the deconfining
transition, $\pardis = 0$ above it. At finite volume $\pardis$ can not vanish
for $\beta > \beta _C$
without vanishing identically, since it is an entire function of $\beta$.
Only in the limit $N_s \to \infty$ singularities develop \cite{Lee}, and
$\pardis$ can vanish.

$M$ is the lowest mass with quantum numbers of a monopole.
In the Landau-Ginzburg model of superconductivity it
corresponds to the Higgs mass. When compared to the inverse penetration depth
of the field, it can give information on the type of superconductor.
We will discuss the determination of $\pardis$ in the next section.

For numerical reasons it will prove convenient to determine \beq
\label{eq:3.24} \rho  =\frac{\diff}{\diff \beta} \log \pardis \ ,
\eeq which, by use of eq.~(\ref{eq:3.17}), amounts to the
difference of the two actions \beq \label{eq:3.25} \rho = \langle
S \rangle _S - \langle S + \Delta S \rangle _{S + \Delta S} \ .
\eeq $\rho$ contains all the informations we need. At finite
temperature the lattice is asymmetric ($N_s^3 \times N_t$ with
$N_t \ll N_s$), the quantities which can be computed are static
and the {\em vev} of a single operator $\mu$, $\pardis$ must be
directly computed. Indeed there is no way of putting a monopole
and an antimonopole at large distance along the $t$ axis as we do
at $T=0$, since at $T\sim T_c$, $N_T a$ is comparable to the
correlation length. $C^{\star}$-periodic boundary conditions in
time ($U_{\mu}(\vec{n},N_t) = U^{\star}_{\mu}(\vec{n},0)$, where
$U^{\star}$ is the complex conjugate of $U$ \cite{Wiese1}) are
needed. The magnetic charge is conserved. If we create a monopole
say at $t=1$, and we propagate it to $t=N_T$ by the changes of
variables described above, the magnetic charge at $t=N_T$ will be
different by one unit from that at $t=0$, and this is inconsistent
with periodic boundary conditions. With $C^*$-periodic boundary
conditions the magnetic field at $N_T$ is opposite to the one at
$t=0$, since under complex conjugation the term proportional to
$\sigma_3$ in eq.(\ref{eulero}) changes sign. By the change of
variables eq.(\ref{eq:3.18}) a magnetic field in then added with
opposite sign at $N_T$.  This produces a dislocation with magnetic
charge $-1$ at the boundary which plays the role of the
antimonopole in eq.(\ref{corr1}).

With a generic choice of the abelian projection different from the
Polyakov line, we can define the operator $\mu$ in a similar way, by sending
\beq
\Pi _{i 0} (\vec{n},t) \to \Pi _{i 0} ^{\prime}(\vec{n},t) \ ,
\eeq
according to eq. (\ref{eq:3.15}).

Again to demonstrate that a monopole is created at $t+1$ we can perform
the change of variables, eq. (\ref{eq:3.18}), and expose a change of the
abelian magnetic field at $t+1$ given by eq. (\ref{eq:3.19}).

However now the resulting change of $\Pi _{0i} (\vec{n},t+1)$ is
\beq
\Pi _{i 0} ^{\prime} (\vec{n},t+1) &=&
U_i(\vec{n},t+1) e^{i A_{\bot i}^M(\vec{n},\vec{y}) \latproj \cdot
\vec{\sigma}}\\
\nonumber
&&
U_0(\vec{n}+ \hat \imath,t+1)\\
\nonumber
&\ &
\left(U_i (\vec{n},t+2)\right)^{\dag}\left(U_0(\vec{n},t+1)\right)^{\dag} \\
\nonumber
&=&  U_i(\vec{n},t+1) U_0(\vec{n}+ \hat \imath,t+1)\\
\nonumber
&&\left((U_0(\vec{n}+ \hat \imath,t+1)\right))^{\dagger}
e^{i A_{\bot i}^M(\vec{n},\vec{y}) \latproj \cdot \vec{\sigma}} \\
\nonumber
&\ & U_0(\vec{n}+ \hat \imath,t+1)\\
\nonumber
&&\left(U_i (\vec{n},t+2)\right)^{\dag}\left(U_0(\vec{n},t+1)\right)^{\dag} \ .
\eeq
The change of $U_i(\vec{n},t+2)$ is by a factor on the right
\beq
\label{eq:3.29}
&&\left(U_0(\vec{n}+ \hat \imath,t+1)\right)^{\dagger}
e^{i A_{\bot i}^M(\vec{n},\vec{y}) \latproj \cdot \vec{\sigma}}\\
\nonumber
&&U_0(\vec{n}+ \hat \imath,t+1) \ .
\eeq
$U_0$ does not commute with $e^{i A_{\bot i}^M(\vec{n},\vec{y}) \latproj
\cdot \vec{\sigma}}$, as it was in the case of $\latbproj$ in the
direction of the Polyakov line.

This looks at first sight as a complication, but it is not. Indeed
the abelian projected phase of a product of links is the sum of
the abelian projected phases of the factors, to ${\cal O}(a^2)$.
From eq. (\ref{eq:3.10}) it follows ${\cal O}(a^2)$ \beq e^{i
\vec{\gamma}^{T1} \cdot \vec{\sigma}} e^{i \gamma_z^1 \sigma_3}
e^{i \vec{\gamma}^{T2} \cdot \vec{\sigma}} e^{i \gamma_z^2
\sigma_3} = e^{i \vec{\Gamma}^T \cdot \vec{\sigma}}{e^{i
(\gamma_z^1 + \gamma_z^2) \sigma_3}} \ . \eeq Hence the abelian
phases of $U_0, U_0^{\dagger}$ in (\ref{eq:3.29}) cancel ${\cal
O}(a^2)$ and the abelian projected field of the modified plaquette
at time $t+2$ is again changed according to eq. (\ref{eq:3.19}).
\section{Numerical results for $\rho$}
\label{results} We will determine the temperature dependence of
$\rho$ on an asymmetric lattice $N_s^3 \times N_t$ ($N_s \gg
N_t$).

For reasons which will be clear in what follows we will
distinguish between abelian projections in which the operator
$\latbproj$ which defines the monopoles is explicitly known, and
projections (like the so called Maximal Abelian) in which the
projection is fixed by a maximizing procedure, and $\latbproj$ is
not explicitly known.

In the first category we studied the following projections. We
will define the operator $\vproj = \proj/|\proj|$ starting from
an operator ${\cal O}$ which is an element of the group, by the
formula
\[ {\cal O} = {\cal O}_0 + i \proj\cdot\vec\sigma\]
\begin{itemize}
\item ${\cal O}$ is connected to the Polyakov line
$L(\vec{n},t) = \Pi _{t^{\prime}=t}^{N_t -1}
U_0(\vec{n},t^{\prime}) \Pi _{t^{\prime}=0}^{t-1}
U_0(\vec{n},t^{\prime})$ as follows\footnote{by $\star$ we
indicate the complex conjugation operation.}: \beq \label{eq:4.29}
{\cal O}(\vec{n},t) = \Pi _{t^{\prime}=t}^{N_t -1}
U_0(\vec{n},t^{\prime}) L^{\star} (\vec{n},0)
\Pi_{t^{\prime}=0}^{t -1} U_0(\vec{n},t^{\prime}) \ ; \eeq
\item ${\cal O}$ is an open plaquette, i.e. a parallel transport on an elementary
square of the lattice \beq \label{eq:4.30} {\cal O}(n) &=& \Pi
_{ij}(n) = U_i(n)U_j(n+\hat{\imath})\\ \nonumber &&
\left(U_i(n+\hat{\jmath})\right) ^{\dag} \left(U_j(n) \right)
^{\dag}\ ; \eeq
\item ``butterfly'' projection, where the projecting operator is
\beq \label{eq:4.31} {\cal O}(n) =  F (n) &=& U_x(n)U_y(n+\hat{x})
\left(U_x(n+\hat{y})\right)^{\dag}\\ \nonumber &&\left(
U_y(n)\right)^{\dag}U_z(n)U_t(n+\hat{z})\\ \nonumber
&&\left(U_z(n+\hat{t})\right)^{\dag} \left(U_t(n)\right) ^{\dag} \
. \eeq
\end{itemize}
The trace of $F$ is the density of topological charge. The
projection defined in (\ref{eq:4.29}) is the Polyakov projection
on a $C^{\star}$-periodic lattice.

From eq.~(\ref{eq:3.24}) and the condition $\langle \mu (\beta=0)
\rangle = 1$ we obtain \beq \label{eq:4.33} \langle \mu (\beta)
\rangle = \exp \left( \int _0^{\beta} \rho (\beta^{\prime}) \diff
\beta ^{\prime} \right) \ . \eeq

If $\pardis$ (defined in any abelian projection) is a disorder
parameter for the deconfining phase transition, we expect that in
the thermodynamic limit ($N_s \to \infty$, $N_t$ constant) $\rho$
goes to a finite bounded value in the strong coupling region, i.e.
in the region below the deconfining transition. In the weak
coupling region $\langle \mu \rangle$ should go to zero in the
same limit, i.e. $\rho$ must go to $- \infty$. In the critical
region we expect an abrupt decrease of $\pardis$, and hence a
negative sharp peak in $\rho$.

A few details about numerical computation. According to eq.
(\ref{eq:3.25}), $\rho$ is the difference between two actions: the
standard $SU(2)$ Wilson action and the ``monopole'' action $S +
\Delta S$.

For the Wilson term simulation can be performed by using an
heat-bath algorithm. This is not possible in the case of the
``monopole'' action. Consider for example the Polyakov projection
and a single monopole operator $\mu(\vec{y},0)$. In the updating
procedure, we can distinguish the following four cases:
\begin{enumerate}
\item update of a spatial link at $t \ne 0,\ 1$. The plaquettes involved have
Wilson's form and the variation of the ``monopole'' action is
linear with respect to the link we are updating;
\item update of a spatial link at $t=0,\ 1$. Although some plaquettes are
modified by the monopole term, the variation of the modified
action $S + \Delta S$ is again linear with respect to the link,
because the field $\Phi$ does not depend on the link we are
changing;
\item update of a temporal link at $t \ne 0$. The local variation of the
action is linear, but the change also induces a change of the
Polyakov loop, i.e. of $\Phi$, according to eq. (\ref{eq:4.29}),
so that there is an effect on the action which is non linear;
\item update of a temporal link at $t=0$. We can not define a force, because
due to the change of the corresponding Polyakov loop the change of
the modified action is non-linear.
\end{enumerate}

In order to perform numerical simulations in the case of the
system described by the ``monopole'' action, an appropriate
algorithm could be Metropolis; however, this method can have long
correlation times. In view of improving decorrelation, we have
performed simulations by using an heat-bath algorithm for the
update of the spatial links and a Metropolis algorithm for the
update of the temporal links.

Similar techniques can be used for the other projections we have
investigated: in all cases, we have chosen to use the heat-bath
updating when the contribution to the action is linear with
respect to the link we are changing and the Metropolis algorithm
when it is not. As a test we verified that the mixed update
correctly works for the Wilson action.

The simulation was done on a 128-node APE Quadrics Machine. We
used an overrelaxed heat-bath algorithm to compute the Wilson term
of eq.~(\ref{eq:3.25}), and a mixed algorithm as described above
for the other term. Far from the critical region at each $\beta$
typically 4000 termalized configurations were produced, each of
them taken after 4 sweeps. The errors are computed with a
Jackknife analysis to the data binned in bunches of different
length. As error we took the maximum of the standard deviation as
a function of the bin length at plateau. In the critical region a
higher statistic is required. Typically the Wilson term is more
noisy. Thermalization was checked by monitoring the action density
and the probability distribution of the trace of the Polyakov
loop.

The discussion of Sect.~\ref{disorder} implies that different
choices for $\vec{A}^M _{\bot}$ are equivalent: eq.
(\ref{eq:3.19}) shows that only the magnetic field of the monopole
determines the value of $\pardis$. In our simulation we used the
Wu-Yang form of $\vec{A}^M _{\bot}$. We checked that the Dirac
form (with different position of the string) gives compatible
results.

In simulations of $S+ \Delta S$ we found that correlation times
are small and under control for $N_t = 4$. For $N_t = 6$ in the
critical region thermalization problems arise and modes with long
correlation time appear. For this reason, we have used mainly
lattices with $N_t = 4$.

\begin{figure}[t]
\begin{center}
\epsfig{figure=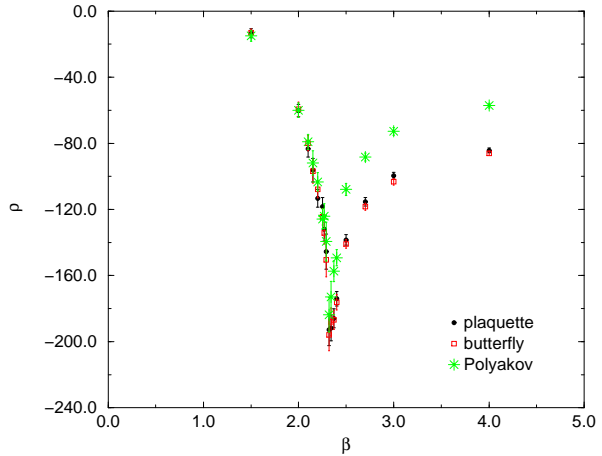, angle=0, width=8.5cm}
\end{center}
\caption{$\rho$ vs. $\beta$ for different abelian projections.
Lattice $12^3 \times 4$.} \label{condtot12.fig} \null\vskip 0.6cm
\end{figure}

Fig.~\ref{condtot12.fig} shows the typical behaviour of $\rho$ for
different abelian projections, for a lattice $12^3 \times 4$. The
negative peak occurs at the expected transition point, $\beta_C$
\cite{Karsch}. Below $\beta_C$ the different projections are
indistinguishable within errors, suggesting that different
monopoles behave in the same way.

\begin{figure}[htb]
\begin{center}
\epsfig{figure=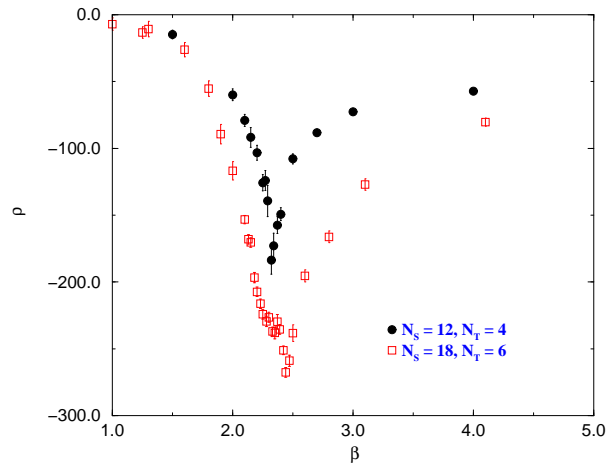, angle=0, width=8cm}
\end{center}
\caption{$\rho$ vs. $\beta$ for different lattice extensions
(lattices $N_s^3 \times N_t$). Polyakov projection.}
\label{confrpoly.fig} \null\vskip 0.3cm
\end{figure}

Fig.~\ref{confrpoly.fig} shows the comparison with a $18^3 \times
6$ lattice. The peak is displaced at the correct $\beta_C$,
showing that it is not an artifact but it is related to
deconfinement of colour \cite{Pieri}.

\begin{figure}[htb]
\begin{center}
\epsfig{figure=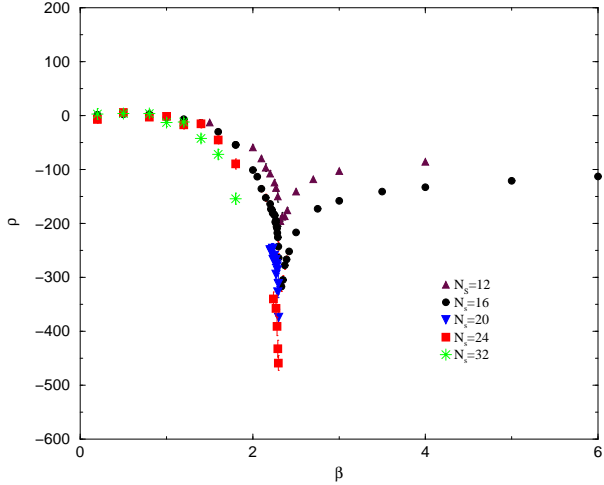, angle=270, width=8cm}
\end{center}
\caption{$\rho$ as a function of $\beta$ for different spatial
sizes at fixed $N_t=4$. Plaquette projection.}
\label{rhopla16.fig} \null\vskip 0.3cm
\end{figure}
\begin{figure}[htb]
\begin{center}
\epsfig{figure=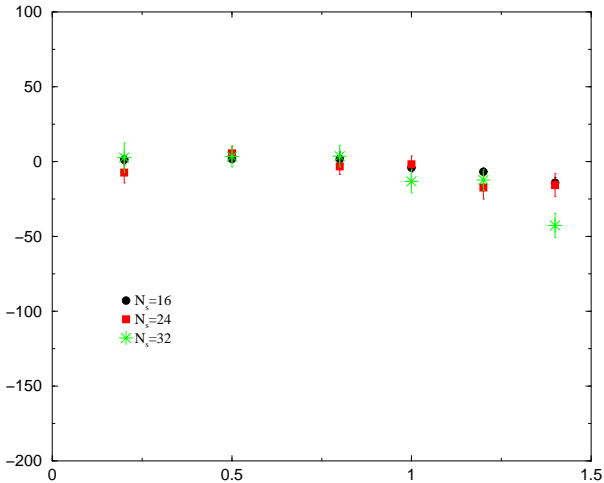, angle=270,width=8cm}
\end{center}
\caption{$\rho$ vs. $\beta$ in the strong coupling region for
lattice sizes $N_s^3 \times 4$. Plaquette projection.}
\label{rhostrong.fig} \null\vskip 0.3cm
\end{figure}

Since different projections give indistinguishable results, for
sake of simplicity we shall only display the plaquette projection
in the following figures.

Fig.~\ref{rhopla16.fig} shows the dependence of $\rho$ on $N_s$ at
fixed $N_t=4$. The qualitative behaviour does not change when we
increase the lattice size. We now try to understand the
thermodynamic limit \cite{CE97,Digiau1}.

In the strong coupling region (cfr. fig.~\ref{rhostrong.fig})
$\rho$ seems to converge to a finite value, which is consistent
with $0$ at low $\beta$'s. Eq. (\ref{eq:4.33}) then implies that
$\pardis \ne 0$ in the infinite volume limit in the confined
phase.

The weak coupling region is perturbative. An estimate of $\rho$ is
the minimum on the ensemble of the configurations $U$ and is given
by the action of classical solutions of the system described by $S
+ \Delta S$: \beq \rho
\mathop{\to}_{\beta \to \infty} &&\! \! \! \! \! \left[
\mathop{\min}_{U} \{ S \} - \mathop{\min}_{U} \{ S+ \Delta S \}
\right] \\ \nonumber &=& - \mathop{\min}_{U} \{ S + \Delta S \} \
, \eeq since $\mathop{\min}_{U} \{ S \} = 0$.

In other systems, where the same shifting procedure has been
applied and studied, this asymptotic value has been analytically
calculated in perturbation theory with the result
\cite{CE97,Digiau1} \beq\label{weakdec} \rho = - c N_s + d \ ,
\eeq where $c$ and $d$ are  constants, i.e $\rho$ goes linearly
with the spatial dimension.

In $SU(2)$ we are unable to perform the same calculation and we
have evaluated the minimum $\mathop{\min}_{U} \{ S + \Delta S \} $
numerically. Some technical remarks on the numerical procedure. An
oversimplified  strategy would be to start from a random
configuration and then decrease  the action by Metropolis-like
steps in which the new configuration is accepted only if its
action is lower. However this procedure will not work, because of
the presence of local minima where often the procedure stops. A
way to overcome this difficulty is to perform an usual Monte-Carlo
simulation where $\beta$ is increased indefinitively during the
simulation \cite{KI83,AL85}. This is equivalent to freeze the
system.

We found useful to integrate the two strategies. Firstly we freeze
the system increasing $\beta$ in the following way:
\begin{enumerate}
\item we thermalize the system at a reasonable $\beta$
(e.g. $\beta = 10$);
\item we increase $\beta$ by a fraction $1/200$ and at the new value of
$\beta$ we perform a number of sweeps (typically 200), looking for
the corresponding minimum of the action;
\item we iterate the step 2 until the minimum of the action looks stable
along a larger number of sweeps (typically 5000).
\end{enumerate}

When this procedure becomes inefficient (typically for $\beta
\approx 10^6$), we go to a Metropolis-like minimization, which is
stopped when the action stays constant within errors.

\begin{figure}[htb]
\begin{center}
\epsfig{figure=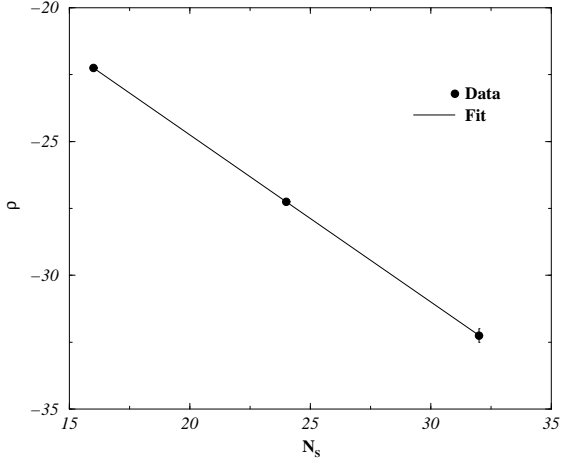, angle=0, width=7.5cm}
\end{center}
\caption{$\rho$ vs. $N_s$ ($N_t=4$) at $\beta = \infty$ in the
plaquette projection. Data are obtained by numerical minimization
of $S + \Delta S$.} \label{rhoweak_su2.fig} \null\vskip 0.3cm
\end{figure}

The result is shown in fig.~\ref{rhoweak_su2.fig} for the
plaquette projection. It is consistent with the linear dependence
of eq. (\ref{weakdec}) with $c \simeq 0.6$ and $d \simeq -12$.
Thus in the weak coupling region in the thermodynamic limit $\rho$
goes to $- \infty$ linearly with the spatial lattice size and \beq
\langle \mu \rangle \mathop{\approx}_{N_s \to \infty} A e^{ (- c
N_s + d )\beta } \to 0,\quad \beta > \beta_C \ . \eeq The magnetic
U(1) symmetry is restored in the deconfined phase.

\begin{figure}[htb]
\begin{center}
\epsfig{figure=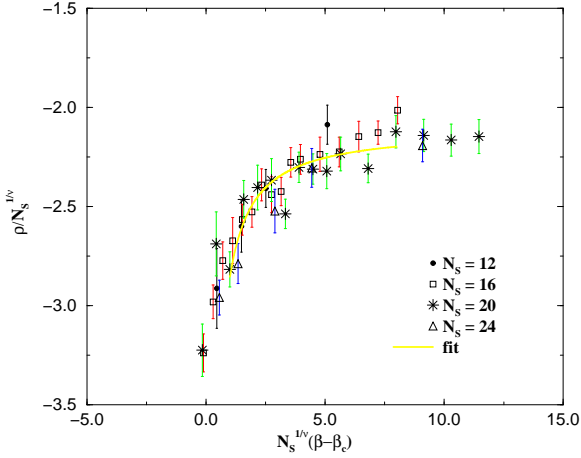, angle=0, width=8.5cm}
\end{center}
\caption{Quality of scaling in the plaquette projection at
$N_t=4$.} \label{scalapla.fig} \null\vskip 0.3cm
\end{figure}

To sum up, $\pardis$ is different from zero at least in a wide
range of $\beta$ below $\beta_C$ and goes to zero exponentially
with the lattice size for $\beta > \beta_C$. The strong coupling
region and the weak coupling one must be connected by a decrease
of $\pardis$; the sharp peak of $\rho$ signals that this decline
is abrupt and takes place in the critical region.

To understand the behaviour of $\rho$ near the critical point we
shall use finite size analysis. By dimensional argument \beq
\pardis = N_s ^{- \delta / \nu}
\Phi \left(\frac{\xi}{N_s},\frac{a}{\xi},\frac{N_t}{N_s} \right) \
, \eeq where $a$ and $\xi$ are respectively the lattice spacing
and the correlation length of the system.

Near the critical point, for $\beta < \beta_C$ \beq \xi \propto
\left( \beta _C - \beta \right)^{- \nu} , \eeq where $\nu$ is the
corresponding critical exponent. In the limit $N_s \gg N_t$  and
for $a / \xi \ll 1$, i.e. sufficiently close to the critical point
we obtain \beq
\pardis = N_s ^{- \delta / \nu}
\Phi \left( N_s^{1/\nu}\left( \beta _C - \beta \right),0,0 \right)
\eeq or equivalently \beq \label{scaling} \frac{\rho}{N_s^{1/\nu}}
= f\left( N_s^{1/\nu} \left( \beta _C - \beta \right)\right) \ .
\eeq The ratio $\rho / N_s^{1/\nu}$ is a universal function of the
scaling variable \beq x =   N_s^{1/\nu} \left( \beta _C - \beta
\right) \ . \eeq Since critical values of $\beta$ and critical
indices of $SU(2)$ pure gauge theory are well known \cite{Karsch},
we can check how well scaling is obeyed by plotting $\rho /
N_s^{1/\nu}$ as a function of $x$.

Fig. \ref{scalapla.fig} shows the quality of the scaling in the
plaquette projection for $\beta _C = 2.2986$ and $\nu = 0.63$.
Similar qualitative results have been obtained for the Polyakov
projection.

As a further check, we can vary $\nu$ and try to estimate ``by
eye'' sensitivity of our data to this exponent. We obtain that in
both projections the scaling relation is satisfied within errors
for $0.57 \leq \nu \leq 0.67$.

In the thermodynamic limit in some region of $\beta$ below the
critical point we expect \beq
\pardis \propto  \left( \beta _C - \beta \right) ^{\delta} \ ,
\eeq which implies \beq \label{rhoas}
\frac{\rho}{N_s^{1/\nu}} = - \frac{\delta}{x}  \ . \eeq

Using eq. (\ref{rhoas}) it should be possible in principle to
determine $\nu$, $\delta$ and $\beta _C$. Our statistic is not
enough accurate to perform such a fit. However, we can determine
$\delta$ using as an input $\beta_C$, $\nu$, which are known, by
parameterizing $\rho$ in a wide range by the form \beq
\frac{\rho}{N_s^{1/\nu}} = -
\frac{\delta}{x}  - c \ , \eeq where $c$ is a constant, as
suggested by fig.~\ref{scalapla.fig}.

Our best fit\footnote{Fits have been performed by using the Minuit
routines.} gives $\delta = 0.24 \pm 0.07$ in the plaquette gauge
and $\delta = 0.12 \pm 0.04$ in the Polyakov gauge. The reduced
$\chi ^2$ is order 1.

This concludes our argument about the thermodynamic limit ($N_s
\to \infty$). The deconfining phase transition can be seen from a
dual point of view as the transition of the vacuum from the dual
superconductivity phase to the dual ordinary phase. That feature
seems to be independent of the abelian projection chosen.
\section{The Maximal Abelian projection}
\label{maxabsection} There are abelian projections which are not
explicitly defined by an operator $\Phi$, but by some
extremization procedure. The prototype is the Maximal Abelian
projection, which is defined by maximizing numerically the
quantity \beq S_U (\{ \Phi \} ) = \sum_{n,\mu} \mbox{tr} \left[
U_\mu(n) \sigma_3
 U^{\dag}_\mu (n) \sigma_3 \right]
\eeq with respect to gauge transformations
\cite{'tHooft3,Laursen}.

The Maximal Abelian projection is very popular since, in the
projected gauge all links are oriented in the abelian direction
within $15 \%$, and therefore all observables are dominated by the
abelian part within $85 \%$. This fact is known as {\em abelian
dominance} \cite{Suzuki}, and could indicate that the abelian
degrees of freedom in this projection are the relevant dynamical
variables at large distances. Moreover, out of the abelian
projected configurations, monopoles seem to dominate observable
quantities ({\em monopole dominance} \cite{Stack}).

With our approach we have a technical difficulty to determine
$\rho$ via $S + \Delta S$ (eq. (\ref{eq:3.25})). At each updating
the operator $\Phi$ and $S + \Delta S$ are only known after
maximization. Accepting or rejecting an updating therefore
requires a maximization, and the procedure takes an extremely long
computer time.

\begin{figure}[tb]
\begin{center}
\epsfig{figure=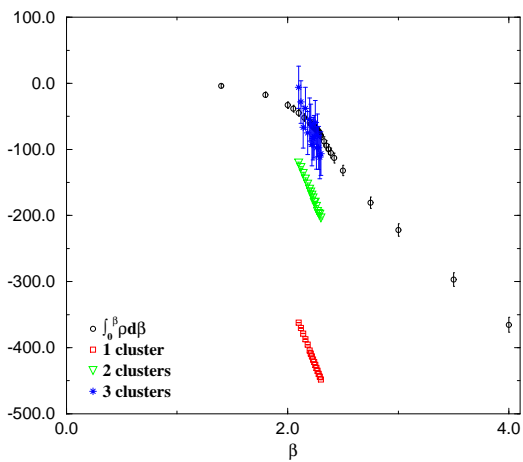, angle=0, width=8cm}
\end{center}
\vskip-0.5cm \caption{$\log \langle \mu \rangle$ reconstruction by
cluster expansion. Plaquette projection, lattice $16^3 \times 4$.}
\label{clus_plaq.fig} \null\vskip 0.3cm
\end{figure}
\begin{figure}[tb]
\begin{center}
\epsfig{figure=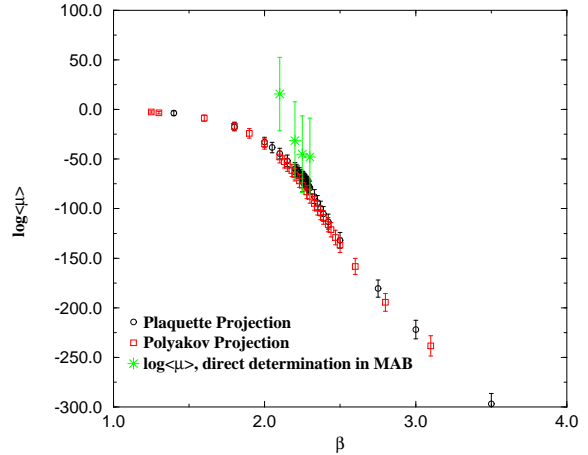, angle=0, width=8cm}
\end{center}
\vskip -0.5cm \caption{$\log \langle \mu \rangle$ in various
abelian projections on a $16^3 \times 4$ lattice.} \null\vskip
0.3cm \label{logmu.fig}
\end{figure}

Therefore in order to study this abelian projection we have to
explore the possibility of measuring $\pardis$ directly, and to
confront with the huge fluctuations coming from the fact that
$\mu$ is the exponential of a sum on a space volume and typically
fluctuates as $\sim e^{N_s^{3/2}}$. We adopt the following
strategy \cite{Digiau1}
\begin{enumerate}
\item we study  the probability distribution of the quantity
\beq \log \mu = - \beta (\Delta S) \ ; \eeq
\item we reconstruct $\langle \mu \rangle$ from the $\log \mu$ distribution
by means of cumulant expansion formula truncated at some order.
\end{enumerate}
This procedure should be compared with that of ref.
\cite{Polikarpov}.

If we have a stochastic variable $X$ distributed with probability
$p(X)$ we have \beq \int d X e^{\beta (X - \langle X \rangle )}
p(X) = e^{\sum_{n \ge 2} \frac{\beta^n}{n !} C_n} \ . \eeq
$\langle X \rangle$ is the mean value and $C_n$ is the $n$-th
cluster. For example, if we call $\Delta = X -\langle X \rangle$
\beq
\begin{array}{l}
C_1 = 0 \ ; \\ C_2 = \langle \Delta^2 \rangle \ ; \\ C_3 = \langle
\Delta^3 \rangle \ ; \\ C_4 = \langle \Delta^4 \rangle - 3
\langle \Delta^2 \rangle  \langle \Delta^2 \rangle \ ; \\ C_5 =
\langle \Delta^5 \rangle - 10 \langle \Delta^3 \rangle \langle
\Delta^2 \rangle \ .
\end{array}\eeq
If $ \log \mu$ were gaussianly distributed with mean value $m$ and
standard deviation $\sigma$, we would have \beq \langle \mu
\rangle = \exp \left( m + \frac{\sigma^2}{2} \right) \ . \eeq

In order to check the method we have explored the cluster
expansion for the $ \log \mu$ distribution in the projection we
have already studied by means of the quantity $\rho$.
Fig.~\ref{clus_plaq.fig} shows a comparison between $\log \langle
\mu \rangle$, taken from the integration of $\rho$ data,  and
cluster expansions truncated at different orders. The first and
the second cluster are insufficient to account for the right
behavior of $\log \langle \mu \rangle$, whereas with the third
cluster added the two determinations are consistent. Moreover the
fourth cluster is zero within statistical errors. It seems that
one can estimate $\langle \mu \rangle $ with a cluster expansion
truncated at the third order. As a rule, the higher clusters are
quite noisy and error bars grow with increasing order. Therefore
this kind of estimation  requires a very high statistics. For this
reason numerical determination of $\log \langle \mu \rangle $ in
the Maximal Abelian projection is possible, but very time
consuming. Our data are displayed in fig.~\ref{logmu.fig}, showing
that monopoles in the Maximal Abelian projection behave in the
same way as monopoles in other projections.

For this kind of simulations, we have used a standard overrelaxed
heat-bath algorithm. For each value of $\beta$ we performed about
50000 measurements, each of them taken after 8 sweeps. In order to
improve the statistics, we have considered eight symmetric
different position of the monopole (namely we have inserted the
monopole at the center of each optant of a cartesian coordinate
system with the origin at the center of the lattice); data
corresponding to each position are analyzed separately with the
method exposed in the previous section and our best value is the
weighted average of the eight measurements. Putting more monopoles
would not improve the statistics, since strong correlations appear
whenever the distance is shorter than the correlation length. Also
these simulations have been performed on a 128-node APE QUADRICS
machine.
\section{Conclusions}
\label{conclusions} We have constructed a disorder parameter
$\pardis$ detecting condensation of monopoles of non abelian gauge
theories defined by different abelian projections. The parameter
is the {\em vev} of an operator which creates a magnetic charge.
$\pardis \ne 0$ signals dual superconductivity. The same
construction has been tested in many known systems
\cite{CE97,Digiau1,Martelli}.

We measure by numerical simulations $\pardis$, or better $\rho =
\frac{\diff}{\diff \beta} \log \pardis$, which contains all the
relevant informations and less severe fluctuations.

An extrapolation to thermodynamic limit (infinite spatial volume)
is possible.

The system behaves as a dual superconductor in the confined phase,
and has a transition to normal at the deconfining phase
transition, where $\pardis \to 0$.

The deconfining $\beta_C$ and the critical index $\nu$ as well as
the critical index $\delta$ describing the way in which $\pardis
\mathop{\to} 0$ when ${T\to T_C}$ can be determined. The first two
quantities are known independently and our determination is
consistent with others. As for $\delta$, defined by \beq
\pardis \mathop\simeq_{T\to T_C} \left( 1- \frac{T}{T_C}\right) ^{\delta} \ ,
\eeq it is $0.20 \pm 0.08$. Different abelian projections
(plaquette, Polyakov, ``butterfly'') give results which agree with
each other.

Our technique proves difficult for the Maximal Abelian projection,
but a direct determination of $\pardis$ looks consistent with
other projections.

In conclusions
\begin{enumerate}
\item Dual superconductivity is at work in the confined phase, and disappears
at the deconfinement phase transition.
\item This statement is independent of the abelian projection defining the
monopoles.
\end{enumerate}

Further theoretical effort is needed to understand the real
symmetry breaking in the deconfined phase, or in the dual
description of {\em QCD}.

Similar results for $SU(3)$ will be presented in the companion
paper.

Finally we stress that, whatever topological excitations are
responsible for colour confinement, counting them is not a right
criterion to detect disorder. Only the {\em vev} of an operator
carrying the appropriate topological charge can be a legitimate
disorder parameter.
\section*{Acknowledgements}
This work is partially supported by EC contract FMRX-CT97-0122 and
by MURST.
\end{document}